\documentclass{PoS}
\usepackage{picins}
\usepackage{graphicx}

\PoS{PoS(LAT2005)310}
\title{Nonperturbative investigation\\ of the diquark potential 
}
\ShortTitle{Nonperturbative investigation of the diquark potential}
\author{Zolt{\'a}n Fodor\\
        E{\"o}tv{\"o}s University Budapest, University of Wuppertal\\
        E-mail: \email{fodor@bodri.elte.hu}}
\author{Christian H{\"o}lbling\\
        University of Wuppertal\\
        E-mail: \email{christian.holbling@cern.ch}}
\author{\speaker{Markus Mechtel}\\
        University of Wuppertal\\
        E-mail: \email{mame@theorie.physik.uni-wuppertal.de}}
\author{K{\'a}lm{\'a}n Szab{\'o}\\
        University of Wuppertal\\
        E-mail: \email{szaboka@general.elte.hu}}

\abstract{We perform an investigation of the static quark-quark-potential both in the confined and the deconfined phase. We discuss conceptual and technical problems and present first results of an exploratory numerical investigation.}
\FullConference{XXIIIrd International Symposium on Lattice Field Theory\\
		 25-30 July 2005\\
		 Trinity College, Dublin, Ireland}

\begin{document}

\section{Introduction}

It is a well known fact, that the Polyakov loop correlator $\left<P(x)P^\dagger(y)\right>$ can be used to measure the static quark-antiquark potential and the correlator of three Polyakov loops tells us about the baryon potential.

As McLerran and Svetitsky showed in \cite{McLerran:1981pb}, the expectation value of every combination of Polyakov loops is related to the free energy of the appropriate configuration of quarks and antiquarks. Thus, it would be interesting to study the correlation of two Polyakov loops in order to determine the diquark potential.

It has been conjectured that diquarks play an important role in phenomenology, e.g. colour superconductivity is motivated by diquark condensates and pentaquarks could be described by using diquarks.

\section{The diquark model}
A single quark transforms as a triplet under SU(3) gauge transformations. Two quarks transform as the tensor product of two triplet states, which can be decomposed as $[3] \otimes [3] = [\bar 3] \oplus [6]$. The antisymmetric $[\bar 3]$-representation transforms in the same way as a single antiquark. This reduction of the effective color charge of the two quarks renders the bound state energetically preferable and leads to an attractive force.

There has been a suggestion of how to measure the quark-quark correlation by Jaffe in \cite{Jaffe:2005md} that has been followed up by other groups \cite{Alexandrou:2005zn} \cite{Orginos:2005vr}. In case of our study we built up our operators out of Polyakov loops only, namly we measured the correlation between two Polyakov loops. This method has the advantage that it is not necessary to fix the gauge. Furthermore we do not need a third quark line like in Jaffe's suggestion. For alternative approaches to measure the free energy see also \cite{Hess:1998sd}, \cite{Babich:2005ay} and \cite{Hubner:2005zj}.

\section{Diquark free energy}
On the lattice, the Polyakov loop expresses the propagation of a single heavy quark along a closed loop in periodic imaginary time: $P(x)=\mathrm{Tr}\left(\prod \limits_{k=1}^{N_t}U_t(x+k\hat t) \right)$.

McLerran and Svetitsky showed \cite{McLerran:1981pb} that the free energy of a configuration of $n$ quarks and $\overline n$ antiquarks at positions $x_1, \dots, x_n$ and $y_1, \dots, y_{\overline n}$ respectively is related to the expectation value of the appropriate combination of Polyakov loops as
\begin{equation}
 \left< P(x_1)\cdots P(x_n) P^{\dagger}(y_1) \cdots P^{\dagger}(y_{\overline n}) \right> = \exp\left[-a N_t (F_{nq, \bar n\bar q} - F_{00}) \right] \, .
 \label{eqn-PLoop-corr-exp}
\end{equation}
Here, $a$ is the lattice spacing, $N_t$ is the lattice extension in the time direction and $F_{00}$ is the free energy of the vacuum without any quarks. Especially for a quark-antiquark pair, this equation leads to the static quark-antiquark potential which is the free energy at zero temperature
\begin{displaymath}
\left< P(x) P^{\dagger}(y) \right> = \exp\left[-a N_t (F_{q\overline q}(|x-y|)-F_{00}) \right] \, .
\end{displaymath}

From equation (\ref{eqn-PLoop-corr-exp}), one can construct the normalized free energy of a specific configuration, i.e. the gain in the free energy of a finite configuration of quarks relative to a configuration where all the quarks are infinitely separated.
\begin{eqnarray*}
  \widehat{F_{nq, \bar n\bar q}} & = & (F_{nq, \bar n\bar q}-F_{00}) - (n+\overline n)\cdot (F_q-F_{00})\\
      & = & -\frac{1}{aN_t}\log\left< P(x_1)\cdots P(x_n) P^{\dagger}(y_1) \cdots P^{\dagger}(y_{\overline n}) \right> + \frac{n+\overline n}{aN_t}\log\left< P(x) \right>\\
      & = & -\frac{1}{aN_t}\log\frac{\left< P(x_1)\cdots P(x_n) P^{\dagger}(y_1) \cdots P^{\dagger}(y_{\overline n}) \right>}{\left< P(x) \right>^{n+\overline n}}\, .
\end{eqnarray*}
Especially for two quarks, this relation leads to the diquark free energy
\begin{displaymath}
 \widehat{F_{qq}} = -\frac{1}{aN_t}\log\frac{\left< P(x) P(y) \right>}{\left< P(x) \right>^2} \, .
\end{displaymath}
The single quark free energy of the vacuum $F_q - F_{00}$ is included into the operator in order to remove the self energy of the two quarks.

\section{The Polyakov loop in pure gauge theory}
\piccaption[]{quenched Polyakov loop density distribution from a $6^4$ lattice in the confinement region \label{Pol-Dist}}
\parpic(6cm,5.8cm)[r][t]{
  \begin{picture}(0,0)(-163,-130)
    \includegraphics[viewport=117 43 237 93, width=5cm]{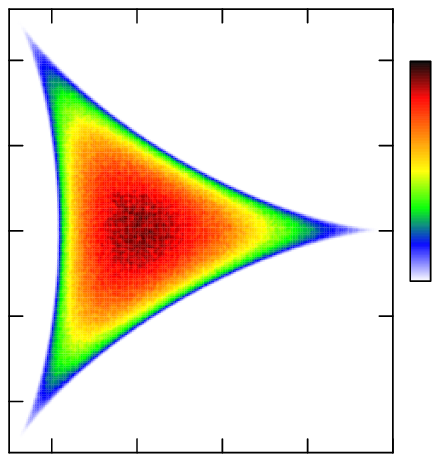}      
  \end{picture}
  \setlength{\unitlength}{0.025bp}
  \begin{picture}(18000,10800)(-100,-2700)
    \scriptsize
    \put(9000,1900){\makebox(0,0){Re P(x)}}
    \put(5700,5675){\rotatebox{90}{\makebox(0,0){Im P(x)}}}
    \put(12350,8400){\makebox(0,0){dense}}
    \put(12350,4800){\makebox(0,0){sparse}}
    \put(7085,2400){\makebox(0,0){-1}}
    \put(8255,2400){\makebox(0,0){ 0}}
    \put(9423,2400){\makebox(0,0){ 1}}
    \put(10593,2400){\makebox(0,0){ 2}}
    \put(6300,8166){\makebox(0,0)[r]{ 2}}
    \put(6300,6996){\makebox(0,0)[r]{ 1}}
    \put(6300,5827){\makebox(0,0)[r]{ 0}}
    \put(6300,4658){\makebox(0,0)[r]{-1}}
    \put(6300,3488){\makebox(0,0)[r]{-2}}
  \end{picture}
}
It is well known, that a pure SU(3) gauge theory exhibits an additional Z(3) center symmetry. The action is not changed if all the time links connecting two neighboring timeslices are multiplied by the same element of Z(3). The Polyakov loop however does change under such a transformation

\vspace{1.5ex}
{\hfill  $P(x) \to \mathrm{e}^{\mathrm{i}\frac{2}{3}\pi n} \cdot P(x) \quad (n=0,1,2)$. \hfill}
\vspace{1.5ex}

This makes the Polyakov loop sensitive to the center symmetry. Thus, in the quenched theory, it serves as an order parameter for the spontaneous breakdown of this symmetry, which is associated with the confinement/deconfinement phase transition.

Since the update algorithm will generate configurations from all three Z(3) sectors with equal probability, the values are distributed symmetrically in all of the three sectors as can be seen in figure \ref{Pol-Dist}.
 Thus the loops of the different Z(3) sectors will cancel out each other, which results in an exactly vanishing expectation value for the Polyakov loop. Therefore, in order to measure a non-vanishing Polyakov loop, one has to break the center symmetry.

There have been some propositions how to break this symmetry. For example, one can take $\left|P(x)\right|$, $\sqrt[3]{\mathrm{Re}[P(x)^3]}$, rotate all values around zero so that the phase of each loop lies in the range from $-\frac{\pi}{3}$ to $\frac{\pi}{3}$ like in \cite{Hubner:2005zj} or just ignore all loops which do not lie in the sector around the positive real axis as has been done in \cite{Nakamura:2004ur}. But all these methods seem somehow arbitrary.
This differs from the case when dynamical fermions are included into the simulation. In the unquenched theory, the fermion determinant explicitly breaks the center symmetry.

\section{Simulation details}
In full QCD, the Polyakov loop does not serve as an order parameter any more, but equation (\ref{eqn-PLoop-corr-exp}) still remains valid. Because of the exponential decay, it is an expensive task to determine the Polyakov loop and its correlators accurately for large lattice extension in the time direction or equivalently for low temperature. But the low temperature region is especially interesting.
One drawback of the dynamical staggered fermions in full QCD is the non-applicability of exponential error reduction such as the Lüscher-Weisz multilevel algorithm \cite{Luscher:2001up}.
Therefore, the lattices which can be studied are restricted to small time extension.

We were able to study lattices with a time extension up to $N_t = 6$ and a spatial extension of $N_s = 18$. The first calculations were done on a $4^4$ and a $6^4$ lattice with the Wilson plaquette action and 4 flavours of staggered fermions with a mass of $m_q a = 0.05$ in lattice units.

\piccaption{$qq$ free energy from a $4^4$ lattice  \label{Fqq(1)}}
\parpic(8cm,4cm)(0.5cm,3.7cm)[r]{
  \includegraphics[width=7cm]{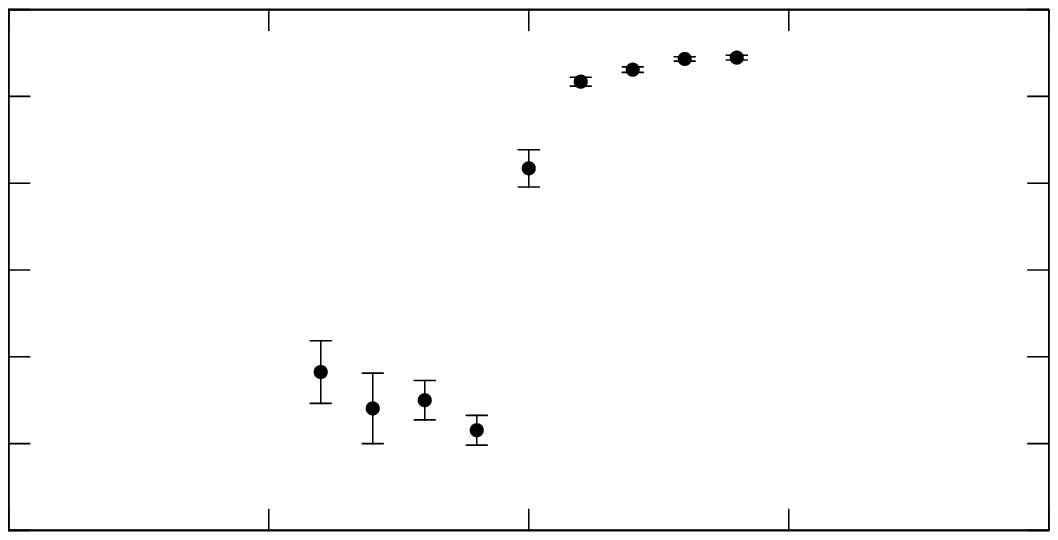}
  \begin{picture}(0,0)
    \footnotesize
    \put(-200,20){\makebox(0,0)[l]{ -0.3}}
    \put(-200,33){\makebox(0,0)[l]{ -0.25}}
    \put(-200,46){\makebox(0,0)[l]{ -0.2}}
    \put(-200,60){\makebox(0,0)[l]{ -0.15}}
    \put(-200,74){\makebox(0,0)[l]{ -0.1}}
    \put(-200,88){\makebox(0,0)[l]{ -0.05}}
    \put(-198,102){\makebox(0,0)[l]{ 0}}
    \put(-175,10){\makebox(0,0){ 4}}
    \put(-135,10){\makebox(0,0){ 4.5}}
    \put(-95,10){\makebox(0,0){ 5}}
    \put(-55,10){\makebox(0,0){ 5.5}}
    \put(-15,10){\makebox(0,0){ 6}}
    \put(-95,0){\makebox(0,0){$\beta$}}
    \put(-210,80){\rotatebox{90}{\makebox(0,0)[r]{$\widehat {F_{qq}}(r=a)$ }}}
  \end{picture}
 }
With this data set, we calculated the diquark free energy at different distances as a function of $\beta$. The result for $r=a$ is shown in figure \ref{Fqq(1)}. As can be seen from this figure, there is a clear correlation between the two Polyakov loops. The sharp jump at $\beta \approx 5$ corresponds to the confinement/de\-con\-fine\-ment phase transition.

\piccaption{$qq$ and $q\overline q$ free energy from a $6^4$ lattice at $\beta = 5.14$, the errors are smaller than the dots. \label{6p4-pot}}
\parpic(8cm,5.5cm)(0.5cm,4.9cm)[r][t]{
  \begin{picture}(0,0)
    \footnotesize
    \includegraphics[width=7cm]{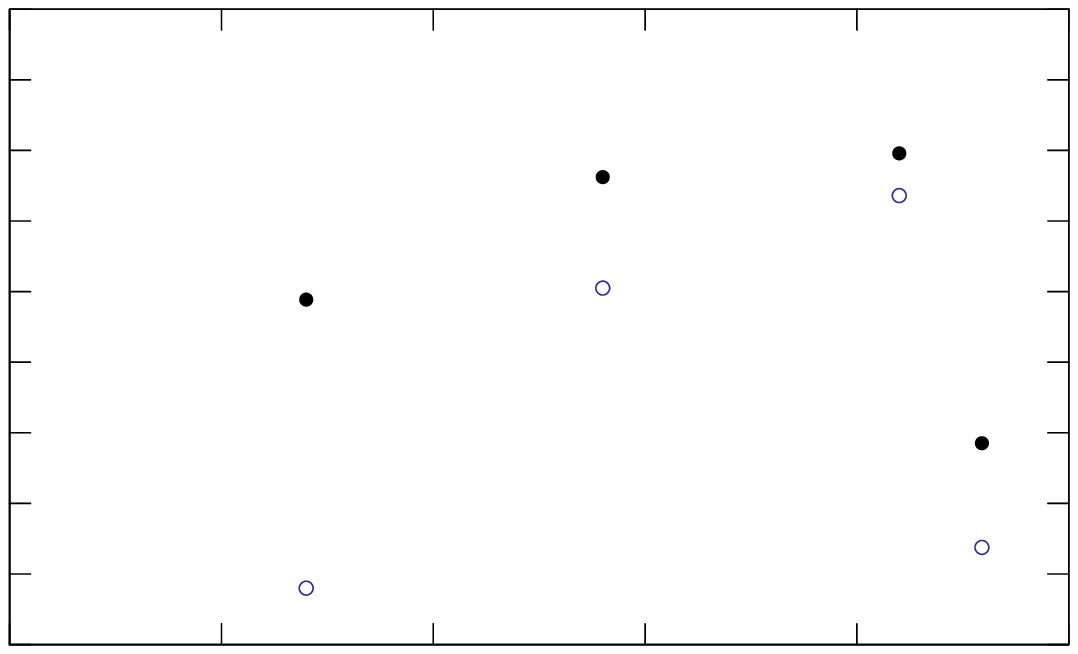}
    \put(-200,23){\makebox(0,0)[l]{ -400}}
    \put(-200,45){\makebox(0,0)[l]{ -300}}
    \put(-200,67){\makebox(0,0)[l]{ -200}}
    \put(-200,89){\makebox(0,0)[l]{ -100}}
    \put(-183,111){\makebox(0,0)[r]{ 0}}
    \put(-210,60){\rotatebox{90}{\makebox(0,0)[c]{ $\widehat F$ $[$MeV$]$}}}
    \put(-180,5){\makebox(0,0){ 0}}
    \put(-145,5){\makebox(0,0){ 0.2}}
    \put(-112,5){\makebox(0,0){ 0.4}}
    \put(-78,5){\makebox(0,0){ 0.6}}
    \put(-45,5){\makebox(0,0){ 0.8}}
    \put(-10,5){\makebox(0,0){ 1}}
    \put(-93,-5){\makebox(0,0){r [fm]}}
    \put(-29,44){\makebox(0,0)[r]{$\widehat {F_{qq}}$}}
    \put(-29,27){\makebox(0,0)[r]{$\widehat {F_{q\overline q}}$}}
  \end{picture}
}
In order to determine the quark-quark correlator for larger distances, we did the same calculations on a $6^4$ lattice at $\beta = 5.14$ which is still in the confinement region and corresponds to a lattice spacing of $a=0.28$ fm. The form of the $qq$ free energy in comparison to the quark-antiquark potential is shown in figure \ref{6p4-pot}. One can see, that the free energy falls off at shorter distances which indicates an attractive force.

It is preferable to determine the form of the $qq$ free energy for larger separations. Since the Polyakov loop decays exponentially with $N_t$, as described by equation (\ref{eqn-PLoop-corr-exp}), we kept the temporal lattice extension fixed to $N_t=6$ and increased the spatial lattice size.

The following results were obtained from a second data set, which was generated using 2+1 staggered fermions with stout smeared links \cite{Morningstar:2003gk} and physical quark masses and the Symanzik tree level improved gauge action on an $18^3\times 6$ lattice. We used the finite temperature configurations which were generated in \cite{Aoki:2005vt}.

\begin{figure}
  \begin{minipage}[b]{0.5\linewidth}
    \centering
    \footnotesize
    \begin{picture}(0,0)(-4,0)
      \includegraphics[width=7cm]{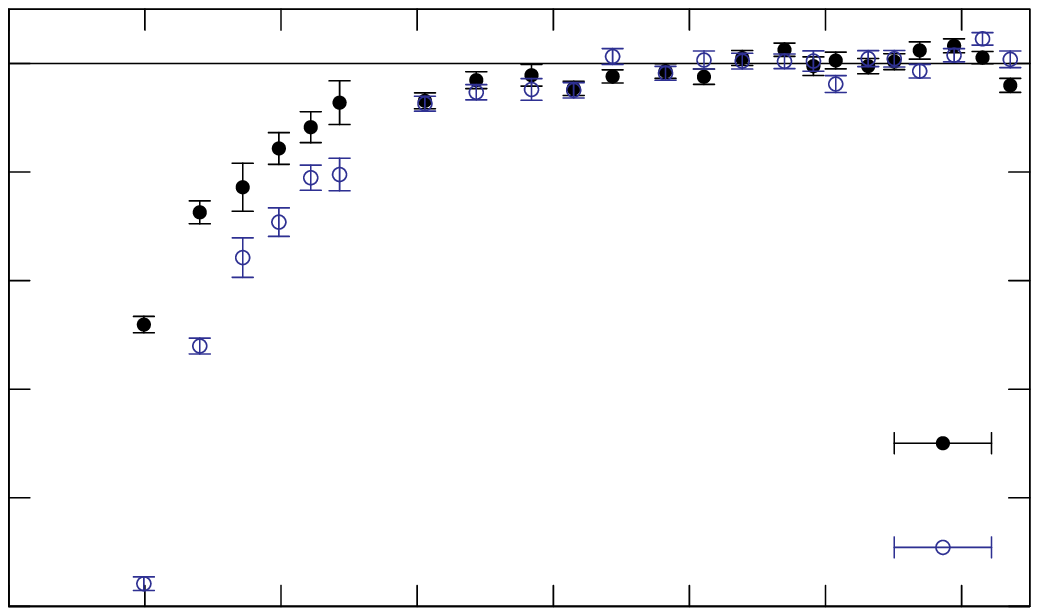}
    \end{picture}
    \setlength{\unitlength}{0.0110bp}
    \begin{picture}(18000,10800)(0,0)
      \put(2200,1650){\makebox(0,0)[r]{-500}}
      \put(2200,3214){\makebox(0,0)[r]{-400}}
      \put(2200,4777){\makebox(0,0)[r]{-300}}
      \put(2200,6341){\makebox(0,0)[r]{-200}}
      \put(2200,7905){\makebox(0,0)[r]{-100}}
      \put(2200,9468){\makebox(0,0)[r]{ 0}}
      \put(2474,1100){\makebox(0,0){ 0}}
      \put(4434,1100){\makebox(0,0){ 0.2}}
      \put(6394,1100){\makebox(0,0){ 0.4}}
      \put(8354,1100){\makebox(0,0){ 0.6}}
      \put(10315,1100){\makebox(0,0){ 0.8}}
      \put(12275,1100){\makebox(0,0){ 1}}
      \put(14235,1100){\makebox(0,0){ 1.2}}
      \put(16195,1100){\makebox(0,0){ 1.4}}
      \put(0,5950){\rotatebox{90}{\makebox(0,0){$\widehat{F}$ [MeV]}}}
      \put(9825,275){\makebox(0,0){r [fm]}}
      \put(14950,4000){\makebox(0,0)[r]{$\widehat{F_{qq}}$}}
      \put(14950,2500){\makebox(0,0)[r]{$\widehat{F_{q \overline{q}}}$}}
    \end{picture}
  \end{minipage}
  \vspace{5mm}
  \begin{minipage}[b]{0.5\linewidth}
    \centering
    \footnotesize
    \begin{picture}(0,0)(-4,0)
      \includegraphics[width=7cm]{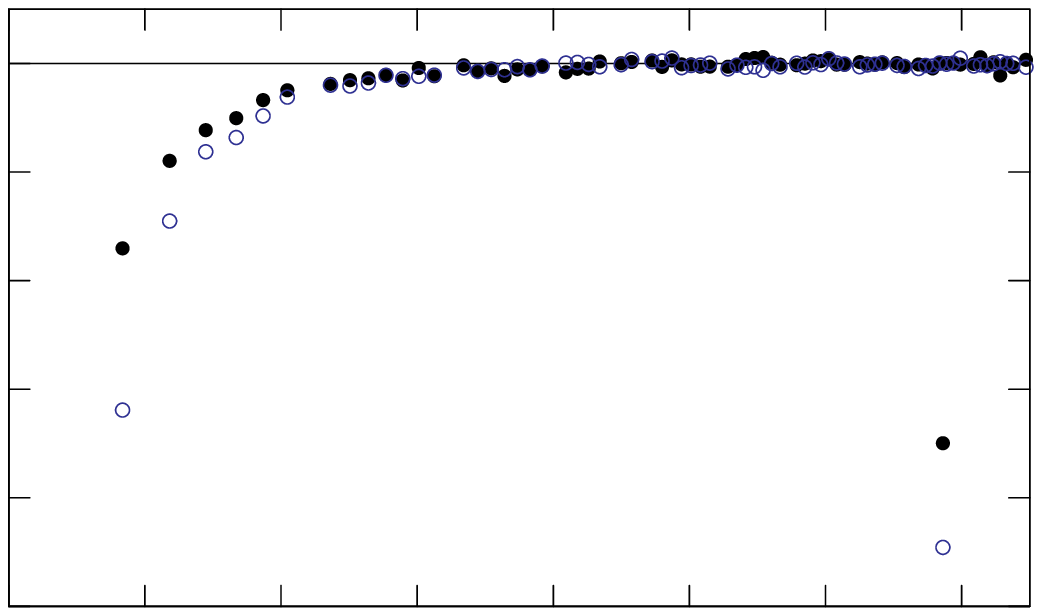}
    \end{picture}
    \setlength{\unitlength}{0.0110bp}
    \begin{picture}(18000,10800)(0,0)
      \put(2200,1650){\makebox(0,0)[r]{-500}}
      \put(2200,3214){\makebox(0,0)[r]{-400}}
      \put(2200,4777){\makebox(0,0)[r]{-300}}
      \put(2200,6341){\makebox(0,0)[r]{-200}}
      \put(2200,7905){\makebox(0,0)[r]{-100}}
      \put(2200,9468){\makebox(0,0)[r]{ 0}}
      \put(2474,1100){\makebox(0,0){ 0}}
      \put(4434,1100){\makebox(0,0){ 0.2}}
      \put(6394,1100){\makebox(0,0){ 0.4}}
      \put(8354,1100){\makebox(0,0){ 0.6}}
      \put(10315,1100){\makebox(0,0){ 0.8}}
      \put(12275,1100){\makebox(0,0){ 1}}
      \put(14235,1100){\makebox(0,0){ 1.2}}
      \put(16195,1100){\makebox(0,0){ 1.4}}
      \put(0,5950){\rotatebox{90}{\makebox(0,0){$\widehat{F}$ [MeV]}}}
      \put(9825,275){\makebox(0,0){r [fm]}}
      \put(15500,4000){\makebox(0,0)[r]{$\widehat{F_{qq}}$}}
      \put(15500,2500){\makebox(0,0)[r]{$\widehat{F_{q \overline{q}}}$}}
    \end{picture}
  \end{minipage}
  \caption{quark-quark free energy compared with the quark-antiquark free energy for \mbox{$\frac{T}{T_c}=0.96$} (left) and $\frac{T}{T_c}=1.14$ (right). In the left plot, a box-averaging has been used for $r>0.5$ fm to reduce the errors. The boxes were smaller than 0.04 fm. In the right plot, the errors are smaller than the dots.}
  \label{18p3x6-fE}
\end{figure}

Figure \ref{18p3x6-fE} shows the diquark free energy as a function of the distance between the Polyakov loops and compare it with the quark-anti\-quark free energy. Note that our data sets correspond to finite temperature samples with $T=0.96 T_c$ and $T=1.14 T_c$. Similar to the quark-anti\-quark attraction, there is an attraction between two quarks. The quark-anti\-quark attraction is significantly stronger than the quark-quark attraction. For our temperatures and for  \mbox{$r > 0.8$ fm} both the $qq$ and the $q\overline q$ signal are compatible with zero.

In agreement with the results of a calculation in the quenched theory presented in \cite{Hubner:2004qf} \cite{Hubner:2005zj}, we see a flattening of the $qq$ free energy above $T_c$ which becomes constant at about the same distance. We also see a flattening below $T_c$, whereas in the quenched calculations, neither the $qq$ nor the $q\overline q$ free energy seem to approach a constant value below $T_c$.

\section{Conclusions}
We see a clear signal for a Polyakov-Polyakov loop correlation in full QCD without any need to fix the gauge. The diquark free energy has a similar form as the quark-antiquark free energy. This is a sign for an attractive force between two quarks, but it is weaker than the attraction between a quark and an antiquark.

Accurately calculating the Polyakov loop expectation value is still an expensive task. In particular, the investigation of zero temperature lattices is restricted by the exponential decay of the Polyakov loop. As a consequence, it is not yet clear, how this Polyakov-Polyakov correlation can be determined in the limit $T \to 0$, which is of physical interest.

\section*{Acknowledgments}
The computations were carried out at E{\"o}tv{\"o}s University on the 330 processor PC cluster of the Institute for Theoretical Physics and the 1024 processor PC cluster of Wuppertal University, using a modified version of the publicly available MILC code \cite{milcLGTcode} and a next-neighbour communication architecture \cite{Fodor:2002zi}.

\bibliographystyle{CasiBibStyle}
\bibliography{References-Proceedings}

\end{document}